\documentclass[twocolumn, notitlepage, 10pt, aps, floatfix, showpacs, prb, citeautoscript]{revtex4-1}
\usepackage{graphicx}
\usepackage{amsmath, amssymb}
\usepackage{bm}
\usepackage{xcolor}
\usepackage[colorlinks, citecolor={blue!50!black}, urlcolor={blue!50!black}, linkcolor={red!50!black}]{hyperref}
\usepackage{bookmark}
\usepackage{microtype}
\usepackage[load=physical,load=abbr, range-units=single]{siunitx}

\newcommand{\pmat}[1]{\begin{pmatrix}#1\end{pmatrix}}
\newcommand{\comment}[1]{}

\begin{document}
\title{Robustness of Majorana bound states in the short-junction limit}
\begin{abstract}
We study the effects of strong coupling between a superconductor and a semiconductor nanowire on the creation of the Majorana bound states, when the quasiparticle dwell time in the normal part of the nanowire is much shorter than the inverse superconducting gap.
This ``short-junction'' limit is relevant for the recent experiments using the epitaxially grown aluminum characterized by a transparent interface with the semiconductor and a small superconducting gap.
We find that the small superconducting gap does not have a strong detrimental effect on the Majorana properties.
Specifically, both the critical magnetic field required for creating a topological phase and the size of the Majorana bound states are independent of the superconducting gap.
The critical magnetic field scales with the wire cross section, while the relative importance of the orbital and Zeeman effects of the magnetic field is controlled by the material parameters only: $g$ factor, effective electron mass, and the semiconductor-superconductor interface transparency.
\end{abstract}
\author{Doru Sticlet}
\author{Bas Nijholt}
\author{Anton Akhmerov}
\affiliation{Kavli Institute of Nanoscience, Delft University of Technology, P.O. Box 4056, 2600 GA Delft, The Netherlands}
\date{24 March 2017}
\maketitle

\section{Introduction}
\comment{Long vs short Josephson junction}
The theory of normal conductor-superconductor (NS) hybrid systems distinguishes two limiting cases: long and short junctions.
In long junctions, the dwell time $\tau_\textrm{dw}$ of a quasiparticle inside the normal region is much larger than the time $\hbar/\Delta$ it spends inside the superconductor (with $\Delta$ the superconducting gap).
In this limit the induced gap inside the semiconductor is equal to $\hbar/\tau_{\textrm{dw}}$, and therefore it varies for different bound states.
In the short-junction or strong-coupling limit, the quasiparticles spend most of their time inside the superconductor, while the normal region effectively acts as a delta function scatterer.
Then in the presence of time-reversal symmetry, the induced gap is close to $\Delta$ for every single Andreev bound state.
In the short-junction limit Andreev bound states have the most weight in the superconductor, and therefore the conventional approach of integrating out the superconductor and obtaining an effective Hamiltonian of the normal system becomes inefficient due to the strong energy dependence of the effective Hamiltonian.

\comment{MBS in short-junction limit}
Systematically studying the short-junction limit is relevant for the creation of Majorana bound states (MBS)~\cite{Qi2011, Leijnse2012, Alicea2012, Beenakker2013, Elliott2015} in semiconductor nanowires\cite{Lutchyn2010, Oreg2010} partially coated with epitaxially grown aluminum that have high interface quality.
These systems were observed to have a well-developed hard induced gap comparable to the gap in the bare Al~\cite{Chang2015}, and subsequently showed zero bias peaks~\cite{Deng2016} and suppressed splitting of low energy states characteristic to MBS~\cite{Albrecht2016}.
The theoretical description of the response of strongly coupled zero-dimensional NS junctions to magnetic field was analyzed in Ref.~\onlinecite{Cole2015}, where the authors report a strong suppression of the effective $g$ factor, potentially leading to the impossibility of inducing a topological phase at magnetic fields below the Clogston limit~\cite{Clogston1962}.

\comment{Use scattering formalism}
Here we extend the analysis of Ref.~\onlinecite{Cole2015} using the scattering formalism that allows us to capture the nonlinear features of the spectrum, and by considering higher-dimensional systems with translational invariance.
The scattering formalism has been routinely applied to short junctions in mesoscopic physics~\cite{[{For review see }] Beenakker2005}.
Relevant works to the present study are on two-dimensional electronic gases with spin-orbit interactions~\cite{Bezuglyi2002, Dimitrova2006}.
However, in Majorana literature the use of the scattering formalism has been limited~\cite{Cheng2012}.
The equivalent of the scattering formalism using the effective Hamiltonian approach amounts to introducing an effective self-energy $\Sigma(E)$ which has a proper dependence on energy $E$~\cite{Liu2012, Stanescu2013, Peng2015} and then neglecting the energy term in the nonlinear eigenvalue problem $[H-\Sigma(E)]\psi = E\psi$, as done in, e.~g.,~Ref.~\onlinecite{Poeyhoenen2016}.

\comment{We find that it is not so bad}
Our overall findings are favorable for the creation of MBS in Al-based NS systems.
Specifically, we find that:
\begin{itemize}
\item The critical magnetic field $B_*$ required to induce a topological phase is independent of the superconducting gap.
 This is valid also beyond the short-junction limit, as long as penetration of magnetic field into the superconductor is negligible.
\item Since $B_*$ is inversely proportional to the wire cross section, the device design can be used to adjust $B_*$ within a broad range.
\item The localization length $\xi_M$ of the MBS does not depend on the superconducting gap, and in optimal conditions it is proportional to the spin-orbit length $l_\textrm{SO}$.
\item Finally, if the interface between the semiconductor and the superconductor has high transparency $T$, then $B_*$ becomes only a slowly varying function of the chemical potential $\mu$, as opposed to its usual oscillatory behavior on the scale of the mode spacing in the nanowire~\cite{Wimmer2010, Lutchyn2011}.
\end{itemize}
Our analytical calculations fully coincide with the results obtained using a numerical scattering approach to short junctions and exact diagonalization of a discretized tight-binding Hamiltonian.
While these conclusions are favorable for the prospect of using weak superconductors for MBS creation, we note that the effects of disorder in the superconductor are not systematically treated here.
Disorder has been recently predicted to have a strong detrimental effect on the creation of MBS in systems that are in the short-junction limit~\cite{Cole2016}.

\comment{Organization of the paper}
The paper is organized as follows.
Section~\ref{sec:formalism} contains a pedagogical review of scattering formalism for the calculation of Andreev spectrum.
The following Sec.~\ref{sec:jjintro} presents scaling arguments supporting our conclusions.
In Sec.~\ref{sec:1stmodel} we compute the dispersion relation of a planar NS junction and discuss the typical device parameters.
Section~\ref{sec:topphases} investigates the Majorana phase diagram and the behavior of the MBS decay length.
In Sec.~\ref{sec:finite} we compare the predictions of Sec.~\ref{sec:topphases} with numerical diagonalization of finite junctions.
Section~\ref{sec:orb} estimates the orbital effect of the magnetic field by computing the Andreev spectrum in a cylindrical geometry in a thin shell limit.
In Sec.~\ref{sec:3d} we confirm our findings using a numerically computed Majorana phase diagram of a three-dimensional model.
Lastly, section~\ref{sec:conc} sums up our conclusions.

\section{Scattering matrix formalism and the short-junction limit}
\label{sec:formalism}
This section reviews the scattering approach to calculating the Andreev bound state spectrum and may be skipped by expert readers.
We start by considering a general NS junction with $n$ superconducting terminals.~\cite{Heck2014}
We use the case $n=1$ in Secs.~\ref{sec:1stmodel},~\ref{sec:topphases},~\ref{sec:3d} and $n=2$ in Sec.~\ref{sec:orb}.

\comment{Andreev bound states condition.}
The levels with $|E|<\Delta$ are Andreev bound states, i.e., coherent superpositions of electron and hole excitations which occur due to Andreev reflections~\cite{Andreev1964} at the interface between the normal region and the superconducting terminals.
The wave function quantization condition on the wave function requires that the total sequence of scattering events results in a phase shift of $2\pi n$.
For the vector of modes $\psi$ incoming from the superconductor to the normal region this condition reads:
\begin{equation}\label{bound}
S_A S_N\psi=\psi.
\end{equation}
Here $S_N$ is the scattering matrix of the normal region, and $S_A$ the scattering matrix of Andreev reflection processes in the superconducting terminals.
The mode vector has electron and hole components $\psi=(\psi_e, \psi_h)$.

\comment{Different bases}
The Andreev reflection matrix assumes a universal form when the superconductor has $s$-wave pairing without any sources of time-reversal symmetry breaking and additionally when the Andreev approximation holds (when the Fermi energy in the superconductor is much larger than $\Delta$).
In the literature, the Andreev spectrum is often calculated in systems where the superconductor Hamiltonian has full spin-rotation invariance (an appropriate approximation for aluminum), making the spin basis a natural choice of basis of $\psi$.
Yet the universal structure of the Andreev reflection matrix does not change in the presence of spin-orbit coupling in the superconductor.
However, in that case it is impossible to choose a spin basis due to lack of spin conservation and it is more appropriate to use a basis where the outgoing modes are time reversed of the incoming modes~\cite{Bardarson2008}.
Throughout the paper we work in the latter basis but explain the relation to the more commonly used spin basis for reference at the end of this section.
Importantly, we neglect the time-reversal symmetry breaking perturbations in the superconductor, restricting ourselves to magnetic fields much lower than critical.

\comment{Normal state scattering matrix.}
The scattering matrix of the normal region is block diagonal in the Nambu space:
\begin{equation}
S_N(E,\mathbf k)=\pmat{S_e(E,\mathbf k) & 0 \\
0 & S_h(E,\mathbf k)},
\end{equation}
where $S_e$ and $S_h$ are the scattering matrices of electrons and holes.
We consider NS junctions with a translational symmetry, and therefore the scattering matrices may depend on the wave vector $\mathbf k$ along the translationally invariant directions.
We choose the hole modes $\psi_e$ as particle-hole partners of the electron modes $\psi_e$.
In this basis the particle-hole symmetry of the scattering matrix reads:
\begin{equation}
\tau_x S^*_N(E,\mathbf k)\tau_x=S_N(-E,-\mathbf k),\label{eq:phs}
\end{equation}
Using the block-diagonal structure of $S_N$ it follows that the normal scattering matrix of holes is the conjugate of the scattering matrix for electrons, at opposite energy and momentum~\cite{Beenakker2015}:
\begin{equation}\label{phs}
S_h(E,\mathbf k)=S_e^*(-E,-\mathbf k).
\end{equation}

\comment{Andreev reflection matrix}
In the same basis, the Andreev reflection matrix reads:
\begin{equation}\label{areflmat}
S_A=\alpha(E)R,\quad R=\pmat{0 & r\\ -r^* & 0},
\quad r=\oplus_j e^{i\phi_j},
\end{equation}
where the index $j$ runs over the terminals, $\phi_j$ is the superconducting phase in lead $j$, and $\alpha(E)=\exp(-i\arccos(E/\Delta))$.~\cite{Beenakker1991}

Following Ref.~\onlinecite{Beenakker1991}, eliminating $\psi$ from Eq.~\eqref{bound}, and using an expression for a block matrix determinant one immediately arrives to a determinantal equation for the bound state energies:
\begin{equation}\label{det}
\det[
1+\alpha^2(E)r^*S_e(E,\mathbf k)rS^*_e(-E,-\mathbf k)
]=0.
\end{equation}
The short-junction limit allows us to further simplify the calculation of the Andreev bound state energies when Thouless energy $E_\textrm{Th} \equiv \hbar/\tau_\textrm{dw} \gg \Delta$.
Thouless energy is the typical energy scale for the matrix elements to change appreciably, therefore in the short-junction limit $S_N(E, \mathbf k) \approx S_N(0, \mathbf k)$ for any $E\lesssim \Delta$.
After replacing $S_e(E)$ with $S_e(0)$, the only energy-dependent term remaining in Eq.~\eqref{bound} is the coefficient $\alpha(E)$.
Since the scattering matrices are invertible, Eq.~\eqref{bound} reads:
\begin{equation}
RS_N\psi
=\alpha^{-1}(E)\psi,\textrm{ or }
S_N^{-1}R^{-1}\psi
=\alpha(E)\psi.
\end{equation}
Adding the two equations yields the following energy eigenproblem:
\begin{equation}
\frac{1}{2}[RS_N+S_N^{-1}R^{-1}]\psi
=\frac{E}{\Delta}\psi.
\end{equation}
Further squaring this equation and using the unitarity of the scattering matrices $S_A$ and $S_N$ we arrive to the eigenproblem expression for the Andreev spectrum:
\begin{equation}\label{nsspec}
\left\{
\frac{1}{2}-\frac{1}{4}
\big[
S_e^\dag(\mathbf k)rS_e^T(-\mathbf k)r^*+\mathrm{H.c.}
\big]
\right\}\psi_e=\frac{E^2}{\Delta^2}\psi_e,
\end{equation}
where the energy argument is suppressed, since $S_e$ is evaluated at $E=0$.
If there is only a single superconducting terminal, the Andreev reflection matrix $r$ reduces to a phase factor, which fully drops out from Eq.~\eqref{nsspec}, as required by gauge invariance.

If the spin is conserved, the above derivation is nearly identical in the spin basis.
The scattering matrices in the spin basis $\tilde S_e$ and $\tilde r$ are related to the basis of time-reversed modes by a transformation
\begin{equation}
\tilde S_{e}=-i\sigma_yS_{e},\quad\tilde r=-i\sigma_y\oplus_j e^{i\phi_j},
\end{equation}
with the Pauli matrices $\bm\sigma$ spin operators and $\sigma_0$ an identity matrix.
The symmetry condition~\eqref{phs}, equation~\eqref{det} for the Andreev spectrum, and eigenproblem for the spectrum in short-junction approximation~\eqref{nsspec} are identical in both bases upon replacing $S_e$ and $r$ with $\tilde S_e$ and $\tilde r$.

\section{Scaling arguments for the MBS properties in the short-junction limit}
\label{sec:jjintro}
\comment{Nothing depends on the superconducting gap}
The superconducting gap enters only as an overall prefactor of the Andreev state energies in Eq.~\eqref{nsspec}, while the specific spectrum depends only on the normal state scattering matrix $S_e$.
This simplification allows us to draw most of our conclusions about MBS properties solely using universal arguments and not by solving a specific model.
For instance, since the Majorana phase transition occurs when Eq.~\eqref{nsspec} has a zero-energy solution, the critical field $B_*$ does not depend on $\Delta$.
This conclusion also extends beyond the short-junction limit, since the zero energy solutions of Eq.~\eqref{nsspec} always coincide with the zero energy solutions of the Eq.~\eqref{bound} and therefore with the full solutions of the Bogoliubov-de-Gennes equation.

Turning to the spatial extent of MBS in the normal region $\xi_M$, we observe that it is limited from below by the coherence length $\xi_S$ in the superconductor.
However $\xi_S$ is often short: For example in aluminum films $\xi_S \sim {100}{\nm}$ due to disorder effects.
If $\xi_M$ is predominantly set by the properties of induced superconductivity in the normal region, $\xi_M$ must also be independent of $\Delta$, since it is a property of the eigenvectors of Eq.~\eqref{nsspec} at $E=0$.

\comment{Thouless energy is the most important energy scale}
If the semiconductor has a cross section $W$, effective electron mass $m_n$, and it is coupled to the superconductor by an interface with transparency $T$, then the Thouless energy equals
\begin{equation}
\label{eq:e_thouless}
E_\textrm{Th} = TN\delta,\quad \delta\equiv\frac{\hbar^2\pi^2}{2m_n (2W)^2},
\end{equation}
where $N$ is the number of transverse bands occupied in the semiconductor, and $\delta$ is the typical interband spacing.
The denominator of the expression for $\delta$ contains $2W$ since it is the total distance traveled by a quasiparticle normal to the interface between two consecutive Andreev reflections.
We focus on the experimentally relevant low-density regime when $N\sim 1$.
In realistic nanowires $m_n\sim 10^{-2}m_e$, and $W\approx \SI{100}{\nm}$, resulting in $\delta \approx \SI{1}{\meV}$, much larger than the superconducting gap in aluminum $\Delta_\textrm{Al} \approx \SI{0.2}{\meV}$, which justifies the short-junction approximation for sufficiently transparent contacts $T\gtrsim 0.1$.

\comment{Thouless energy sets the energy scale for $E_Z$}
In order for the spectral gap to close in the normal region---or, in other words, for a topological phase transition to appear---the scattering matrix $S_e$ must change by $\mathcal{O}(1)$ since, in the presence of time-reversal symmetry, all the Andreev bound states have the same energy $E=\Delta$.
For the Zeeman field to cause such a perturbation, the electron spin must precess by a large angle during the propagation inside the scattering region.
This results in a condition $E_Z \tau_\textrm{dw}/\hbar\sim 1$, or equivalently
\begin{equation}
B_* \sim E_\textrm{Th}/g \mu_B,\label{eq:b_critical}
\end{equation}
with $g$ the effective gyromagnetic factor and $\mu_B$ the Bohr magneton.

\comment{Orbital field has the same scaling with the geometric parameters of the device}
The orbital effect of the magnetic field causes an additional time-reversal symmetry breaking perturbation to the normal scattering region.
It becomes significant [causes an $\mathcal{O}(1)$ change of $S_e$] when the flux penetrating the scattering region becomes comparable to the flux quantum $\Phi_0=h/e$.
This defines another scale of the magnetic field, characterizing the importance of its orbital effect $B_\textrm{orb}\sim h/eW^2$.
Comparing $B_\textrm{orb}$ with $B_*$ we get
\begin{equation}
\label{eq:relative_field_strength}
\frac{B_\textrm{orb}}{B_*} \sim \frac{g\mu_Bm_n}{e\hbar TN}.
\end{equation}
If transparency is high and the number of modes is low, then the relative strength of the orbital and Zeeman effects of the magnetic field is a material parameter dependent on the $g$ factor and the effective mass.
For realistic materials this factor is $\mathcal{O}(1)$, which is in line with our results in Secs.~\ref{sec:orb} and~\ref{sec:3d}.

\comment{MBS size is given by $l_\textrm{SO}$.}
Turning to the spatial extent of the MBS $\xi_M$, we observe that it must diverge in the topological regime in the absence of spin-orbit coupling.
The spectral gap at a finite momentum appears already in the first order perturbation in spin-orbit strength $\alpha$, and hence $\xi_M \sim \alpha^{-1}$.
Finally, in the optimally tuned situation $N\sim 1$, and $B \sim B_*$, so that $S_N$ only depends on two energy scales: $\delta$ and the spin-orbit energy $E_\textrm{SO} = m_n \alpha^2/2\hbar^2$.
This means that there is only a single length scale inversely proportional to $\alpha$, the spin-orbit length $l_\textrm{SO} = \hbar/m_n\alpha$, and hence $\xi_M\sim l_\textrm{SO}$.

\comment{In an open junction $\tau_\textrm{dw}$ is a smooth function, so there are no oscillations in $B_*$.}
The scattering approach highlights another important property of the Majorana phase diagram, the relation between $T$ and the oscillatory behavior of $B_*$.
If $T\sim 1$, there is little scattering at the NS interface, and $\tau_\textrm{dw}$ becomes a smooth function of the chemical potential $\mu$.
Combining this with Eq.~\eqref{eq:b_critical} we conclude that $B_*$ must also depend on $\mu$ in a smooth fashion.
In the opposite limit $T\ll 1$, $E_\textrm{Th}$ reduces on resonance, when $\mu$ matches the bottom of a subband in the semiconducting region.
Away from the resonance, when there are no available states at the selected energy, $E_\textrm{Th}$ becomes very large.
This behavior of Thouless energy results in the appearance of a sharp minimum in $B_*$ whenever $\mu$ matches the bottom of a new band in the semiconductor region.
In Sec.~\ref{sec:1stmodel} we confirm the relation between the interface transparency and the oscillatory nature of the Majorana phase boundary.

\comment{The long junction behavior is not so different}
These findings are different from the predictions of a purely 1D phenomenological model~\cite{Lutchyn2010, Oreg2010} with the Hamiltonian
\begin{equation}
\label{eq:1D_majo_ham}
H_\textrm{1D} = \left(\frac{p^2}{2m_n} -\mu +\frac{\alpha}{\hbar} \sigma_y p\right)\tau_z + \Delta' \tau_x + E_Z \sigma_z,
\end{equation}
where the induced superconductivity enters as a phenomenological pairing term $\Delta'$, and the momentum $p$ is limited to a direction along the nanowire.
The induced gap follows from a perturbation theory in the weak coupling limit between the semiconductor and the superconductor.
Therefore the phenomenological model is not directly applicable to the strong coupling regime for highly transparent junctions.
The Hamiltonian $H_\textrm{1D}$ undergoes a topological phase transition when $E_Z^2 = \Delta'^2 + \mu^2$, and therefore $B_*$ explicitly depends on $\Delta'$.
This difference, however, is due to the shortcomings of the effective model, and in reality our conclusions also hold in the weak-coupling/long junction limit.
In the long junction limit $\Delta'\approx E_\textrm{Th}$, immediately leading us to the conclusion that $B_*$ and $\xi_M$ are independent of the intrinsic superconducting gap $\Delta$.
If a long junction is transparent $T\sim 1$, then the Fermi momentum drops out of the level quantization condition, hence resulting in the lack of oscillations of $B_*$ as a function of $\mu$.
Finally, the rest of our conclusions follow in a similar fashion for the long junctions from the dimensional analysis of Eq.~\eqref{eq:1D_majo_ham} after the identification $\Delta'\sim E_\textrm{Th}$.

\section{Model}
\label{sec:1stmodel}
\subsection{General solution}
\label{sub:gen}
\comment{Clean infinite superconductor limit is not unreasonable.}
To verify our general arguments, we consider a specific model, a semiconductor nanowire in contact with a large superconductor.
We consider the effective superconductor thickness to be infinite, unlike the typical experimental situation where the superconductor thickness is around \SI{10}{\nm}.
This limit is nevertheless a reasonable approximation due to the large Fermi surface size mismatch between the superconductor and the semiconductor.
A much larger Fermi surface in the superconductor means that most electron trajectories approaching the nanowire from the superconductor side must be reflected back.
Full internal reflection in combination with diffuse scattering allows the superconductor to accommodate quasiparticle trajectories much longer than the superconducting coherence length $\xi_\mathrm S=\hbar v_{s,F}/\Delta$, making the superconductor effectively infinite.

\begin{figure}[t]
\includegraphics[width=0.8\columnwidth]{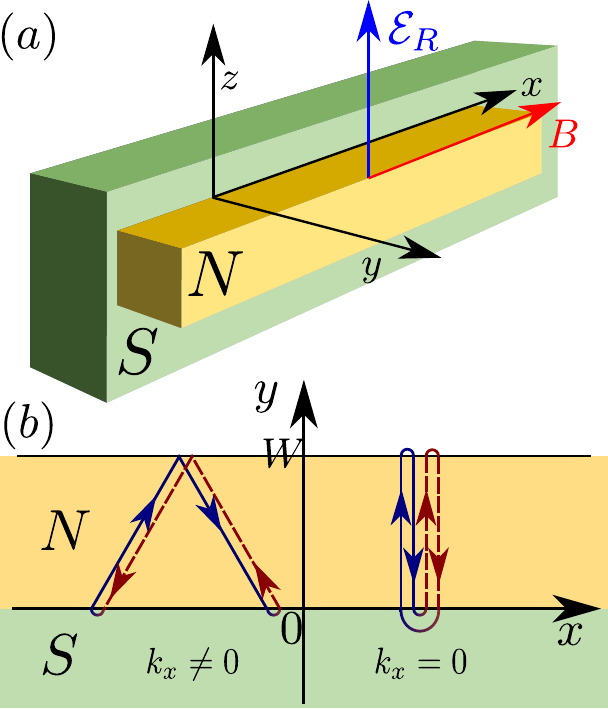}
\caption{(Color online) (a) Nanowire with width $W$ oriented along the $x$ direction and coupled to a bulk superconductor.
The magnetic field is parallel to the wire, while the Rashba electric field points along the $z$ direction.
(b) Semiclassical bound state trajectory in the two-dimensional nanowire. Electrons (solid blue) and holes (dashed red) specularly reflect at the boundary with vacuum and undergo Andreev reflection at the interface with the superconductor.
Two types of bound states are shown: at finite longitudinal momentum $k_x$ (left) and vanishing momentum $k_x=0$ (right).
The induced gap may only close at $k_x = 0$.}
\label{fig:1stmodel}
\end{figure}

\comment{We consider a planar geometry without orbital effect of the magnetic field}
We first consider the device geometry shown in Fig.~\ref{fig:1stmodel}.
The nanowire is oriented along the $x$ axis, NS interface is at $y=0$, and the outer boundary of the wire is at $y=W$.
The superconductor occupies the half-space $y<0$.
Neglecting the orbital effect of the magnetic field makes the motion in $z$ direction separable and reduces the problem to a purely two-dimensional geometry, shown in Fig.~\ref{fig:1stmodel}(b).

\comment{We use Rashba 2DEG + BdG}
The normal state Hamiltonian of the system is that of a Rashba two-dimensional electron gas coupled to a material with negligible spin-orbit interaction and Zeeman coupling~\cite{BenDaniel1966}:
\begin{equation}\label{danduke}
H=\bigg[\bm p\frac{1}{2m(y)}\bm p-\mu(y)\bigg]\sigma_0
+\frac{1}{2\hbar}\{\alpha(y),\bm\sigma\times\bm p\}\cdot\hat z
+E_Z(y)\sigma_x.
\end{equation}
Here $\bm{\sigma}$ are the spin Pauli matrices, and $\bm p$ the momentum operator.
The chemical potential $\mu$ and effective mass $m$ are
\begin{equation}
\label{eq:potentials}
\mu(y)=
\begin{cases}
\mu_{n},& y \in(0,W)\\
\mu_s,& y < 0
\end{cases},\,
m(y)=
\begin{cases}
m_{n},& y \in(0,W) \\
m_s,& y < 0
\end{cases}.
\end{equation}
Additionally, we neglect the spin-orbit coupling and the magnetic field effect in the superconductor, therefore restricting the model to $B\ll B_c$, with $B_c$ the critical field of the superconductor:
\begin{equation}
\label{eq:spin_terms}
\alpha(y) = \alpha \Theta(y)\Theta(W-y),\quad E_Z = \frac{g\mu_B B}{2}\Theta(y)\Theta(W-y),
\end{equation}
with $\Theta$ the Heaviside step function and $\alpha$, the Rashba spin-orbit coupling strength.
The spin-orbit term in Eq.~\eqref{danduke} is symmetrized using anticommutators to ensure current probability conservation at the interface.
The Zeeman energy $E_Z$ is due to a magnetic field of magnitude $B$ oriented along the wire direction.
The effective electric field generating the Rashba spin-orbit coupling is $\bm{\mathcal E}_R= 2m_n\alpha /\hbar g\mu_B\hat z$.
To compare the short-junction approximation with exact diagonalization results, we use the Bogoliubov-de Gennes (BdG) Hamiltonian:
\begin{equation}\label{sc}
H_\textrm{BdG}=\begin{pmatrix}H(B) & \Delta(y) \\ \Delta(y) & -H(-B)\end{pmatrix},
\end{equation}
with $\Delta(y) = \Delta \Theta(-y)$.
The choice of a step-function pairing potential is justified due to a density of states mismatch between the superconductor and semiconductor by more than $10^6$, which renders the self-consistency condition on $\Delta$ unimportant.

To make further analytical progress, we neglect the spin-orbit coupling in the $y$ direction $\alpha\sigma_x p_y$.
This is a valid simplification since in semiconductor nanowires the spin-orbit length $l_\mathrm{SO}$ is usually larger than the nanowire width $W$.
We later verify the validity of this approximation by computing the exact expression for the topological phase boundary and by including the transverse spin-orbit coupling in all the tight-binding simulations.

The wave function $\psi_n(k_x, y)$ in the nanowire satisfies the boundary condition $\psi_n(k_x, W) = 0$ and has the general form
\begin{equation}
\psi_n=u_+ c_+\sin[k_+(W-y)]+u_-c_-\sin[k_-(W-y)],\label{eq:psi_n}
\end{equation}
with
\begin{eqnarray}\label{spinors}
u_+&=&\frac{1}{\sqrt2}\pmat{1\\e^{i\varphi}},\quad
u_-=\frac{1}{\sqrt2}\pmat{e^{-i\varphi}\\-1},\\
e^{i\varphi}&=&\frac{E_Z-i\alpha k_x}{\sqrt{E_Z^2+\alpha^2k_x^2}},\notag
\end{eqnarray}
and $c_\pm$ unknown amplitudes.
The wave function in the superconducting lead has the form
\begin{equation}
\psi_s=
\begin{pmatrix}
  a_{\textrm{in},\uparrow} e^{iqy} + a_{\textrm{out},\uparrow} e^{-iqy}\\
  a_{\textrm{in},\downarrow} e^{iqy} - a_{\textrm{out},\downarrow} e^{-iqy}\\
\end{pmatrix}\label{eq:psi_s},
\end{equation}
with $a_{\textrm{in}}$ the amplitudes of the incoming modes, $a_\textrm{out}$ the amplitudes of the outgoing modes, and the relative signs chosen to ensure that the incoming and outgoing modes are time-reversed of each other.
Finally, the momenta $q$ and $k_\pm$ normal to the NS interface are fixed by the dispersion relation at energy $E$:
\begin{eqnarray}\label{momenta}
q &=& \left[\frac{2m_s}{\hbar^2}(E+\mu_s)
-k_x^2\right]^{1/2},\notag\\
k_\pm &=& \left[\frac{2m_n}{\hbar^2}(E+\mu_n\mp\sqrt{E_Z^2+\alpha^2k_x^2})
-k_x^2
\right]^{1/2}.
\end{eqnarray}

We use the wave function continuity at $y=0$ as well as the current conservation condition on the wave function derivative normal to the interface, in the $y$ direction:
\begin{equation}\label{bc}
m_n^{-1}\psi'_n(k_x, 0) = m_s^{-1}\psi'_s(k_x, 0).
\end{equation}
Solving for $c_\pm$ and $a_\textrm{out}$ for given $a_\textrm{in}$ we obtain the scattering matrix:
\begin{equation}\label{trsscattmat}
S_e=\frac{1}{2}
\pmat{
(r_+-r_-)e^{i\varphi} & r_++r_- \\
-r_+-r_- & (r_--r_+)e^{-i\varphi}
},
\end{equation}
with the reflection phases of different spin projections given by
\begin{equation}\label{scattphase}
r_\pm=\frac{v_s-iv_\pm\cot(k_\pm W)}{v_s+iv_\pm\cot(k_\pm W)}.
\end{equation}
Here we introduced the transverse velocities $v_s=\hbar q/m_s$ in the superconductor lead and $v_\pm=\hbar k_\pm/m_n$ in the nanowire.

The scattering matrix holds generally at energies below the superconducting gap $|E/\Delta|<1$.
In the short-junction approximation, $S_e$ is evaluated at Fermi energy $E=0$.
Then the Andreev bound spectrum follows immediately upon solving the eigenvalue problem~\eqref{nsspec}:
\begin{equation}\label{andreev}
E=\pm\Delta\sqrt{1-\frac{1}{4}|r_+(k_x)-r_-(k_x)|^2\cos^2(\varphi)}.
\end{equation}
This dispersion relation admits no zero energy solutions for $k_x\ne 0$ and $\alpha \ne 0$.
The parameters values yielding $E(k_x=0) = 0$ are the topological phase transitions, and they occur when
\begin{equation}\label{toptrans}
(r_++r_-)|_{k_x=0}=0.
\end{equation}

In the derivation of the Andreev spectrum~\eqref{andreev} we neglected the effect of spin-orbit interactions in the $y$ direction since $W \ll l_\mathrm{SO}$.
For completeness, we analyze the impact of this spin-orbit coupling on the condition for gap closing at $k_x=0$.
In the presence of transverse spin-orbit coupling, one needs to take into account the Hamiltonian~\eqref{danduke} including the $\alpha\sigma_xp_y$ term.
Then the boundary condition at the NS interface needs to be modified in order to ensure the current conservation.
Integrating the Schr\"odinger equation near the interface $y=0$ yields:
\begin{equation}
\frac{1}{m(y)}p_y\sigma_0\psi(y)\bigg|_{0^-}^{0^+}
+\frac{\alpha}{\hbar}\sigma_x\psi(0)=0.
\end{equation}
Since at $k_x = 0$, the Hamiltonian~\eqref{danduke} commutes with $\sigma_x$, the scattering states in the semiconductor region are also eigenstates of $\sigma_x$.
Matching the wave functions at the NS interface and solving the scattering problem results in a condition for closing the excitation gap identical to Eq.~\eqref{toptrans}, but with the modified scattering phases $r_\pm$~\eqref{scattphase}:
\begin{equation}\label{kx_0}
k_\pm=\bigg[
\frac{2m_n}{\hbar^2}(E+\mu_n+E_\textrm{SO}\mp E_Z)
\bigg]^{1/2},\quad
E_\textrm{SO}=\frac{m_n\alpha^2}{2\hbar^2}.
\end{equation}
For our parameter choice $E_\textrm{SO}\approx\SI{40}{\mu \eV}$, and is more than two orders of magnitude smaller than $\delta$.
This confirms that spin-orbit dynamics in $y$ direction is negligible.

\subsection{Typical physical parameters of the heterostructure}
\label{sec:phys_params}
To consider a specific example of the system parameters we take InSb nanowires~\cite{Mourik2012} with effective mass $m_n=0.015\,m_e$ (here $m_e$ is the free electron mass), and with spin-orbit length $l_\mathrm{SO}=\hbar^2/m_n\alpha\approx \SI{250}{\nm}$.
The superconductor in the heterojunction is aluminum, with $m_s \approx m_e$ and chemical potential $\mu_s=\SI{11.7}{eV}$.
A thin Al film has the bulk superconducting gap $\Delta=\SI{0.25}{\meV}$ and the critical magnetic field $B_c$ that varies from around \SI{1.5}{T} to \SI{2}{T}.

While most of our results scale trivially with $W$, we choose $W=\SI{100}{\nm}$ whenever it is necessary to compare the magnetic field or chemical potential scales to the experimental parameters.
This results in $\delta \approx \SI{0.6}{\meV} \gg \Delta$, well within the requirements of the short-junction approximation.

\subsection{Modeling the NS interface}
\label{sub:interf}

\comment{Interface transparency is the main parameter that we want to simulate correctly, and it is set by the velocity ratio in the model}
The final crucial parameter of the hybrid system is the transparency of the NS interface.
In the model Hamiltonian~\eqref{danduke} the interface properties are set by the velocity ratio $v_\pm/v_s$, the only way the superconductor Hamiltonian parameters enter the scattering matrix~\eqref{scattphase}.
We use the $v_s$ as a free parameter allowing us to study the effect of the interface properties on the topological phase diagram.

\comment{A transparent interface is not unreasonable to expect due to the charge accumulation}
While the Fermi energy difference between aluminum and the semiconductor may span several orders of magnitude, the Fermi velocities do not differ so much because of a smaller effective mass in narrow band semiconductors.
Specifically, the Fermi velocity in aluminum is $v_s \sim \SI{2e6}{\m/\s}$, while $v_\pm \sim \SI{2e5}{\m/\s}$ at a relatively low $\mu_n = \SI{3}{\meV}$, resulting in $T\gtrsim 0.4$.
In real systems, the microscopic interface properties such as coupling strength and charge accumulation further influence the interface transparency.
In the absence of a Schottky barrier, extra charge density at the interface smoothens the sharp change in velocity between the semiconductor and the superconductor and further enhances the transparency.

\comment{Long junction allows to determine transparency, while the observed short junction behavior only puts a lower bound on it}
Transparency of the NS interface is hard to measure experimentally due to the complicated geometry of the normal metal-nanowire-superconductor samples.
The experiments using high $\Delta$ superconductors such as NbTiN are in a long junction regime allowing us to estimate $T$ because the induced superconducting gap is $\approx T \delta$.
On the other hand, tunneling spectroscopy only provides a lower bound on the transparency: $T \gtrsim \Delta/\delta$ in the short-junction regime.

\comment{Therefore we check what happens with a fully transparent interface as well as semi-reflective interface}
To explore the impact of interface transparency on the MBS properties we adopt two choices of $v_s$: the highly transparent interface corresponding to $v_s \approx v_\pm$ and an interface with a finite transparency where we fix the value of $\mu_s$ to a constant.
For convenience we choose an anisotropic mass in the superconductor $m_{s,y} = m_n$, $m_{s,x} = m_\parallel \gg m_n$, so that $\mu_s = \mu_n$ results in a perfect transmission at a $k_x=0$ and $B = 0$.
The condition $m_\parallel \gg m_n$ ensures that $v_s$ only weakly depends on $k_x$, as it should due to the Fermi surface in the superconductor being much larger than in the semiconductor.
In most calculations we use $m_\parallel = 10\,m_n$, however our conclusions are not sensitive to this choice (see Appendix~\ref{sec:app} for details).

Adapting the calculations of Sec.~\ref{sub:gen} to the case of anisotropic mass and transparent limit yields the same result for the excitation spectrum~\eqref{andreev}, up to replacing the transverse momentum $q$ in the superconductor with
\begin{equation}\label{aniso}
q = \left[\frac{2m_n}{\hbar^2}\mu_s
-\frac{m_n}{m_\parallel}k_x^2\right]^{1/2}.
\end{equation}

\begin{figure}[t]
\includegraphics[width=\columnwidth]{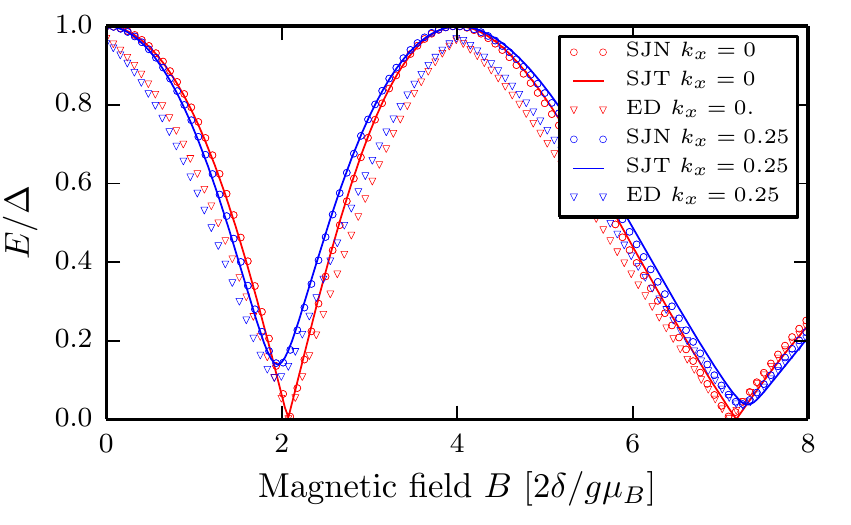}
\caption{(Color online) Comparison of the Andreev spectra of a transparent NS junction between analytical short-junction predictions (SJT), numerical short-junction results including the spin-orbit interaction in the $y$ direction (SJN), and exact diagonalization of the BdG Hamiltonian~\eqref{sc} (ED).
The system parameters are chosen as in Sec.~\ref{sec:phys_params} and \ref{sub:interf}.
The longitudinal momentum is either $k_x=0$ (red), or finite such that the spectrum stays always gapped (blue).
Momentum $k_x$ is in units $k_F^0=\sqrt{2m_n\mu_n/\hbar^2}$, and $\mu_s=\mu_n=\SI{3}{\meV}$.}
\label{fig:spec_comp}
\end{figure}

\subsection{Comparison with tight-binding dispersion simulations}
To verify the correctness of the spectrum in the short-junction limit, Eq.~\eqref{andreev}, we compare the analytical expressions with dispersion relations calculated using a Hamiltonian~\eqref{danduke} discretized on a square lattice with lattice constant $a=\SI{0.5}{\nm}$ and simulated using Kwant package\cite{Groth2014}.
We first numerically obtain the scattering matrix $S_e(k_x)$ of the normal region and use it as an input to Eq.~\eqref{nsspec} to obtain the dispersion relation of the hybrid system.

A further comparison is provided by modeling the hybrid system using the full BdG Hamiltonian~\eqref{sc} and calculating several eigenstates closest to the Fermi level.
In this case, the junction remains infinite along the wire, but instead of a superconducting lead, we attach a large superconductor with width $W_\mathrm{SC} \approx \SI{9}{\mu\m} \gg \xi_\mathrm{S}$.

A comparison between analytics and the two numerical methods at a fixed chemical potential is shown in Fig.~\ref{fig:spec_comp} and shows nearly perfect agreement between different methods.
Slight deviations of exact diagonalization results occur near the bulk gap, caused by corrections to the short-junction approximation.

\section{Analysis of the topological phase diagram}
\label{sec:topphases}

\subsection{Phases boundaries and the spectral gap}
\label{sec:phas-bound-spectr}

\begin{figure}
\includegraphics[width=\columnwidth]{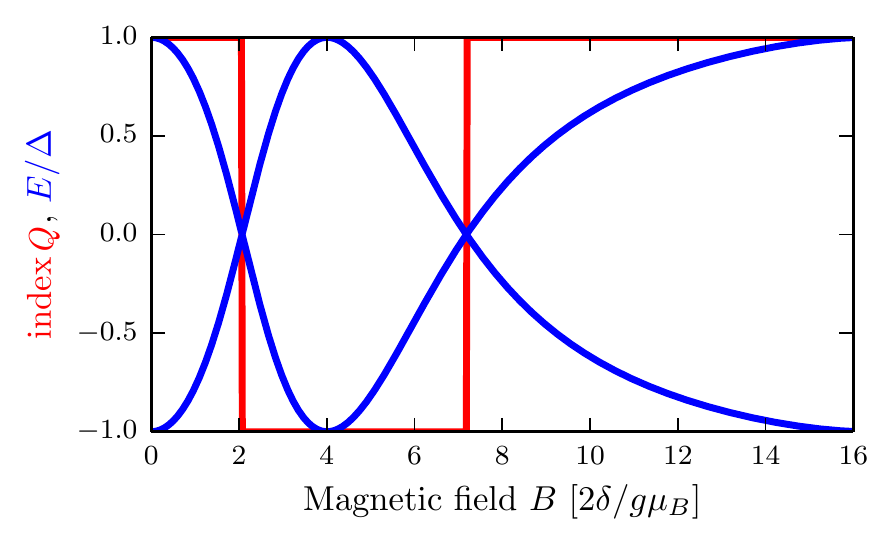}
\caption{(Color online) The Andreev spectrum and the topological index $\mathcal Q$~\eqref{index} as a function of magnetic field.
Trivial phase has $\mathcal{Q}=+1$, topological phase $\mathcal{Q}=-1$.
The junction is in the transparent regime with $\mu_s=\mu_n=\SI{3}{\meV}$.}
\label{fig:index}
\end{figure}

\comment{We compute topological invariant of this model}
Equation~\eqref{toptrans} yields the closing of the spectral gap and the topological transitions in the model.
This allows us to define the topological invariant of the Hamiltonian as
\begin{equation}\label{index}
\mathcal Q=\mathrm{sign}[\mathrm{Im}(\sqrt{-r_-^*r_+})]|_{k_x=0},
\end{equation}
where the sign of the square root is fixed by the analytic continuation and chosen such that $\mathcal Q=1$ in the trivial state.
A typical spectrum at $k_x = 0$ as well as $\mathcal Q$ for a fixed $\mu$ is shown in Fig.~\ref{fig:index}.

\begin{figure}[tb]
\centering
\includegraphics[width=\linewidth]{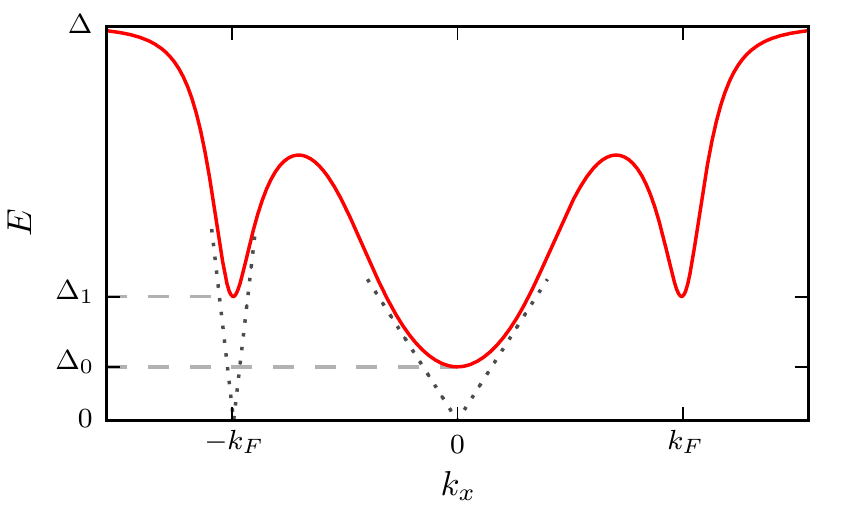}
\caption{A schematic of the dispersion relation of the junction for typical parameters.
The dispersion relation near local minima $\Delta_0$ and $\Delta_1$ of the Andreev state energy is well approximated with a gapped Dirac dispersion relation, with Dirac cones marked with dashed lines.
When the number of available modes in the semiconductor increases, the number of minima at finite momenta grows, but the outermost minimum stays located approximately at $k_x = k_{n,F}$.}
\label{fig:dispersion}
\end{figure}

\comment{We calculate the phase diagram from the spectrum}
In addition to identifying the topological phase boundaries for each set of parameters $(B, \mu_n)$ we calculate the spectral gap
\begin{equation}\label{specgap}
\Delta_\mathrm{spec}=\min_{k_x}|E(k_x)|,
\end{equation}
with $E(k_x)$ given by Eq.~\eqref{andreev}. The minimization is carried over all $k_x$ present in the superconductor.
In general, the dispersion relation has several local minima, as shown in Fig.~\ref{fig:dispersion}, with the total number of minima approximately equal to the number of transverse modes in the normal region.

\begin{figure}
\includegraphics[width=\columnwidth]{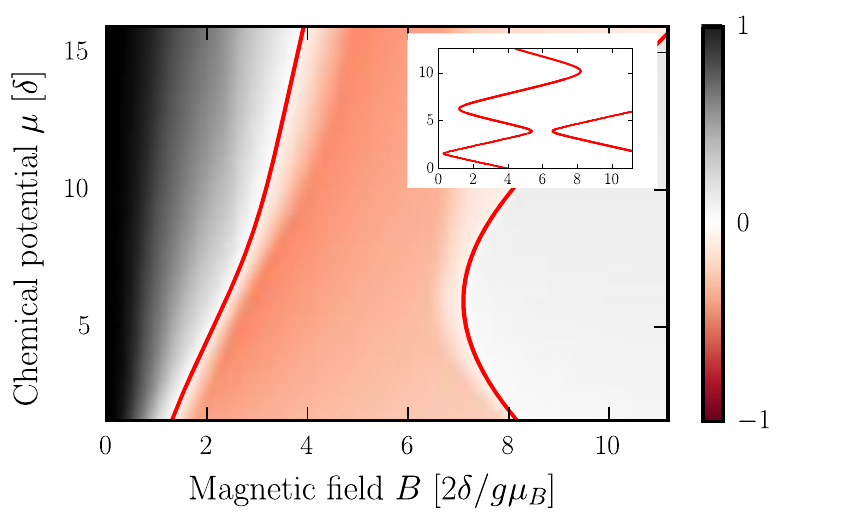}
\caption{(Color online) The spectral gap times the topological index $\mathcal Q \Delta_\mathrm{spec} /\Delta \in(-1,1)$ as a function of chemical potential $\mu$ and magnetic field $B$.
Here we consider a transparent NS interface $\mu=\mu_s=\mu_n$ and an anisotropic mass in the superconductor $m_\parallel=10\;m_n$, $m_\perp = m_n$.
The phase boundaries Eq.~\eqref{toptrans} are given by continuous red lines.
The central region of the phase diagram is the topological phase $\mathcal Q=-1$.
The inset shows the phase boundaries in a similar parameter range for a junction with $\mu_s=\SI{11.7}{eV}$, $m_s=0.015\;m_n$ resulting in a low interface transparency.
}
\label{fig:gaps}
\end{figure}

\comment{It is reasonable overall, and has all the usual asymptotes.}
The resulting topological phase diagram of a transparent junction (with $\mu_s = \mu_n$) is shown in Fig.~\ref{fig:gaps}.
For typical junction parameters (as described in Sec.~\ref{sec:phys_params}) $g\mu_B B_c \approx 9\,\delta$, and the phase diagram for higher field values does not apply to such junctions.
The minimal value of the critical field in this phase diagram corresponds to $B_*\approx \SI{0.7}{\tesla}$.
Near the topological phase transitions $\Delta_\textrm{spec} = \Delta_0$, the spectral gap at $k_x=0$ (see Fig.~\ref{fig:dispersion}), and it varies linearly with the distance $\varepsilon$ from the phase transition either along the $\mu$ or $E_Z$ axis:
\begin{equation}
  \label{eq:dirac_point_gap}
  \Delta_\textrm{spec} = \Delta_0 \sim \Delta \frac{\varepsilon}{\delta}.
\end{equation}
Deep in the topological phase, $\Delta_\textrm{spec}$ is limited by the gap $\Delta_1$ at $k_x \approx k_{n,F}$ (see Fig.~\ref{fig:dispersion}), similar to the phenomenological model of Eq.~\eqref{eq:1D_majo_ham}.
Since $\Delta_\textrm{spec}$ must vanish linearly with $\alpha$ in this regime, we get
\begin{equation}
  \label{eq:SOC_gap}
  \Delta_\textrm{spec} = \Delta_1 \sim \Delta \sqrt{\frac{E_{\textrm{SO}}}{\delta}}.
\end{equation}
In both estimates we assumed $\mu \sim E_Z \sim \delta$, and $T\sim 1$.
Comparing Eqs.~\eqref{eq:dirac_point_gap} and \eqref{eq:SOC_gap} we find the energy scale for the transition between the two behaviors $\varepsilon_* \sim \sqrt{E_\textrm{SO} \delta}$.

\comment{But the phase boundary has no wiggles due to transparency.}
The most unusual feature of the topological phase diagram in Fig.~\ref{fig:gaps} is the smooth behavior of the topological phase boundary, different from the hyperbolically-shaped boundary $E_Z^2 > \Delta'^2 + \mu^2$ of the phenomenological models\cite{Lutchyn2010,Oreg2010,Lutchyn2011}.
This difference appears not due to the short-junction limit---since the magnitude of the gap does not impact the topological phase boundary---but rather because of the high interface transparency.
The inset in Fig.~\ref{fig:gaps} shows the shape of the topological phase boundary where we have reduced the transparency by fixing $\mu_s$ at a high value.
We find that in this case the phase boundary has a hyperbolic shape predicted by the phenomenological model.

\subsection{Decay length of MBS}

\comment{We calculate the decay length in the Dirac limit}
Exact evaluation of the MBS decay length starting from Eq.~\eqref{andreev} is not possible because the spectrum in the short-junction approximation does not correspond to a local Hamiltonian (the same fact manifests in the complex nonlinear dispersion of the Andreev states).
Nevertheless, the decay length is approximated well by assessing the contributions of different local minima of the dispersion relation, as shown in Fig.~\ref{fig:dispersion}.
A gapped Dirac cone with velocity $v$ and gap $\Delta$ results in a wave function decay length $\xi = \hbar v/\Delta$ at $E=0$.
The size of the MBS $\xi_M$ is set by the slowest decaying component of the wave function, or the largest $\xi$.

\comment{Using scaling arguments we find that the crossover between two regimes happens at $\sim E_{SO}$}
Once again, it is instructive to estimate $\xi$ using scaling arguments in two regimes: near a topological transition and deep in the topological phase.
At the phase transition point the slope of the Dirac cone at $k_x=0$, $v_0 \propto \alpha$ since without spin-orbit coupling the band touching at $k_x = 0$ must have a parabolic shape.
Since the bulk superconductor gap $\Delta$ must enter the spectrum only as an overall prefactor, we get
\begin{equation}\label{eq:v_k_0}
\hbar v_0 \sim \frac{\Delta\, W^2}{ l_\textrm{SO}},
\quad \xi_0 = \frac{\hbar v_0}{\Delta_0} \sim \frac{W^2 \delta}{l_\textrm{SO}\varepsilon}.
\end{equation}
The velocity at the outermost Dirac point must not depend on $\alpha$, resulting in
\begin{equation}\label{eq:v_k_f}
\hbar v_1 \sim \Delta\, W,
\quad \xi_1 = \frac{\hbar v_1}{\Delta_1} \sim l_\textrm{SO}.
\end{equation}
The two length scales $\xi_0$ and $\xi_1$ become equal at $\varepsilon \sim E_\textrm{SO} \ll \sqrt{E_\textrm{SO} \delta} = \varepsilon_*$.

\begin{figure}[t]
\includegraphics[width=\columnwidth]{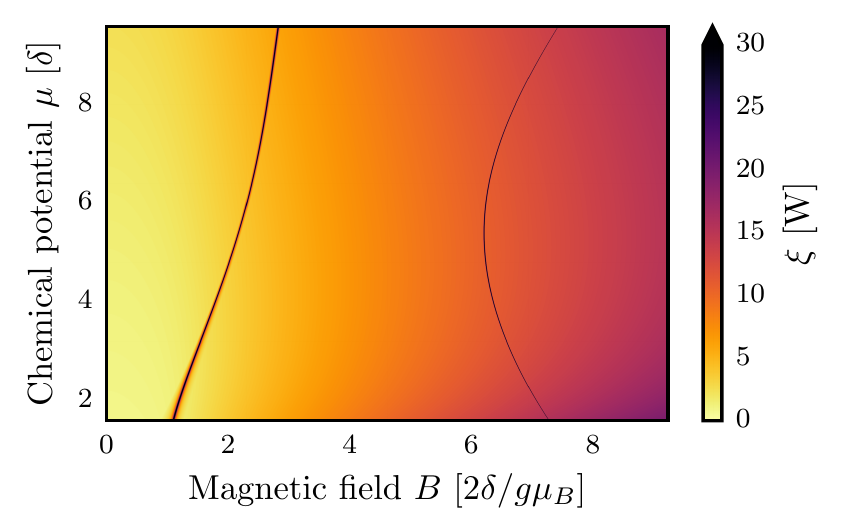}
\caption{(Color online) The largest decay length of subgap modes in units of wire width $W$, as a function of chemical potential $\mu$ and magnetic field $B$.
At the topological transitions the decay length rapidly diverges.
The junction is in a fully transparent regime with $\mu_n=\mu_s$ and $m_\parallel =\infty$ in the superconductor.
}
\label{fig:decay_diag}
\end{figure}

\comment{The small crossover energy scale is clearly visible in the numerical calculation of $\xi$.}
We obtain the behavior of $\xi$ in the tight-binding simulations using Kwant~\cite{Groth2014} for the same parameters as in Fig.~\ref{fig:gaps}.
In order for the self-energy to become local in the $x$ coordinate we neglect the transverse dispersion in the superconductor and set $m_\parallel = \infty$.
We then integrate out the superconductor and add a self-energy to the semiconductor.
Finally, similar to Ref.~\onlinecite{Nijholt2016} we perform an eigendecomposition of the translation operator in the $x$ direction at zero energy to obtain the evanescent waves $\psi\propto e^{-\kappa x}$, with $\kappa$ the eigenvalue of the translation operator.
The largest decay length is:
\begin{equation}\label{ximnum}
\xi = \max\mathrm{Re}[\kappa]^{-1},
\end{equation}
where the maximum is taken over all the eigenvalues.
Then in the topological phase $\xi_M = \xi$ .
The results are presented in Fig.~\ref{fig:decay_diag}.
The divergence in decay lengths seen in Fig.~\ref{fig:decay_diag} corresponds to topological transitions identical to the ones found in Fig.~\ref{fig:gaps}.
Figure~\ref{fig:decay_diag} also confirms that $\xi_M$ saturates at a distance $\varepsilon \sim E_\textrm{SO}$ away from that phase transition (here $E_\textrm{SO} \approx\SI{40}{\mu eV}$).

We now refine the scaling arguments of Eqs.~\eqref{eq:v_k_0} and \eqref{eq:v_k_f} by using Eq.~\eqref{andreev}.
In particular, near the topological transition, the decay length is determined by the spectral gap $\Delta_0$ and the velocity $\hbar v_0=|\partial E/\partial k_x|_{\Delta_0=0,k_x=0}$:
\begin{equation}
\Delta_0=\frac{\Delta}{2}|r_++r_-|_{k_x=0},
\end{equation}
with Fermi velocity
\begin{equation}
\hbar v_0=\frac{\alpha\Delta}{E_Z}.
\end{equation}
Therefore the MBS decay length $\xi_M$ is inversely proportional to the magnetic field and spin-orbit length near the Majorana phase transitions.

\begin{figure}[t]
\includegraphics[width=\columnwidth]{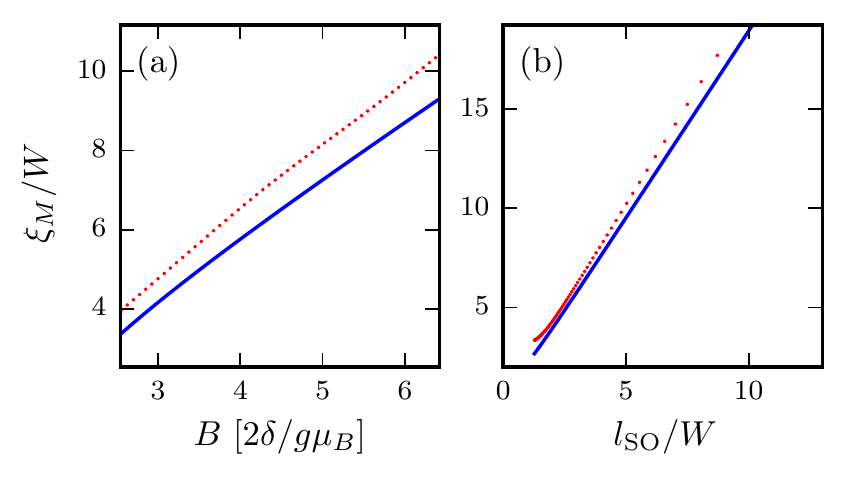}
\caption{
(Color online)
Scaling behavior of the MBS decay length $\xi_M$ deep in the topological phase.
Comparison between symbolic calculation of the decay length from the linearization of the energy dispersion~\eqref{andreev} (blue lines) and the numerical calculation of the slowest decaying mode Eq.~\eqref{ximnum} (red markers).
(a) The decay length has a linear dependence with magnetic field $B$ deep in the topological region, $\mu=\SI{3}{\meV}$.
(b) Linear dependence with spin-orbit length $l_\mathrm{SO}=\hbar^2/m\alpha$.
The magnetic field is $B=\SI{1.5}{T}$ and chemical potential $\mu=\SI{3}{\meV}$.}
\label{fig:compAB}
\end{figure}

Deep in the topological phase it is more difficult to obtain a closed form approximation for the decay length.
Instead, we find the Fermi momentum and the spectral gap by performing numerical minimization of the energy dispersion~\eqref{andreev}.
The Fermi velocity near $k_F\ne 0$ follows immediately:
\begin{equation}
\hbar v_1=\frac{\Delta}{2}\frac{\partial}{\partial k_x}|r_++r_-|_{\alpha=0,k_x=k_F}.
\end{equation}
Taking the ratio~\eqref{eq:v_k_f}, it follows that the MBS decay length does indeed grow linearly with magnetic field and spin-orbit length deep in the topological phase [see Fig.~\ref{fig:compAB}], in qualitative agreement with the numerical calculation using Eq.~\eqref{ximnum}.

Our results for the scaling of $\xi_M$ with $B$ and $l_\textrm{SO}$ agree with the predictions of the phenomenological 1D model both near the topological transition or deep in the topological phase (see, e.~g.,~Ref.~\onlinecite{Klinovaja2012}), but we find no dependence of $\xi_M$ on $\Delta$.

\section{Spectrum of finite length junctions}
\label{sec:finite}

\begin{figure}[t]
\includegraphics[width=\columnwidth]{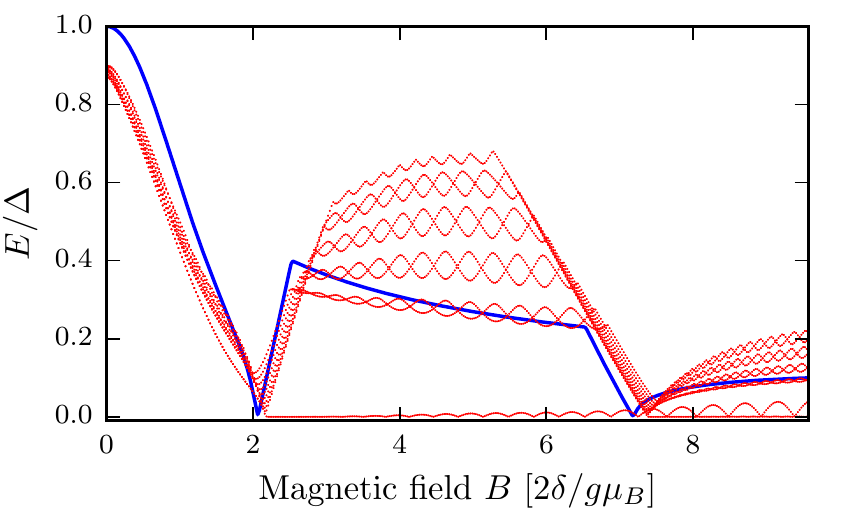}
\caption{(Color online) Comparison between the predictions of the analytical short-junction approximation and a numerical spectrum of a finite NS junction.
Solid line: $\Delta_\textrm{spec}$ calculated using Eqs.~\eqref{andreev} and~\eqref{specgap} as a function of magnetic field.
Dotted lines: 10 lowest energy states in a finite size NS junction using the same parameters.
The magnetic field is in units of $2\delta/g\mu_B$, with level spacing $\delta$ defined in Eq.~\eqref{eq:e_thouless}.
The junction is in a transparent regime, the superconductor has anisotropic mass $m_\parallel = 10\, m_n$, and the rest of parameters are as specified in Sec.~\ref{sec:phys_params}.
At a single end of the nanowire, there is only one MBS in the topological phase ($\SI{1}{T}\lesssim B \lesssim \SI{3}{T}$, for semiconductor width $W=\SI{100}{\nm}$) and two in the trivial phase at high field, due to the chiral symmetry.}
\label{fig:finite}
\end{figure}

To directly verify the existence of a MBS and the applicability of the short-junction limit to our system, we solve the discretized BdG Hamiltonian of a large rectangular system with a finite superconductor.
The system is divided into semiconductor and superconductor regions, both modeled by the BdG Hamiltonian~\eqref{sc}.
The length of the system is $L=\SI{3}{\mu \m}$, sufficiently long to ensure that the overlap between MBS is small.
Further, we choose the width of the superconductor sufficiently large $W_\mathrm{SC}=\SI{1.4}{\mu\m} \approx 2\,\xi_\mathrm S$.
The lattice constant in the tight-binding simulation is \SI{10}{\nm}.
Finally, the remaining model parameters are chosen according to Sec.~\ref{sec:phys_params} and Sec.~\ref{sub:interf}.
We determine numerically several lowest energy states and compare them with $\Delta_\textrm{spec}$ calculated in Sec.~\ref{sec:topphases}, as shown in Fig.~\ref{fig:finite}.

We observe that the energy of most subgap states is bounded from below by an energy slightly lower than $\Delta_\textrm{spec}$, as expected close to the short-junction regime.
At $B > B_* \approx \SI{1}{\tesla}$ the system enters a topological regime and states with $E \ll \Delta_\textrm{spec}$ formed by two coupled MBS appear.
The coupling of these states decays exponentially with the size of the nanowire $L$.
Finally after the system undergoes the second gap closing and enters the trivial phase at $B \approx \SI{3}{\tesla}$ additional low energy states appear due to the presence of chiral symmetry of the Hamiltonian~\eqref{danduke}~\cite{Schnyder2009, Ryu2010, Tewari2012}.
We therefore conclude that our calculations fully apply to finite nanowires in the short-junction regime.

Furthermore, we verify in Appendix~\ref{sec:app2} using exact diagonalization that the critical field $B_*$ is indeed independent on the superconducting gap $\Delta$.
With increasing $\Delta$ above $E_\mathrm{Th}\approx \SI{1}{\meV}$ the system exits the short-junction regime. Nevertheless, the critical magnetic field stays constant, in agreement with the proof of Sec.~\ref{sec:jjintro}.

\section{Orbital field effect in thin shell approximation}
\label{sec:orb}

\begin{figure}[tb]
\centering{\includegraphics[width=\columnwidth]{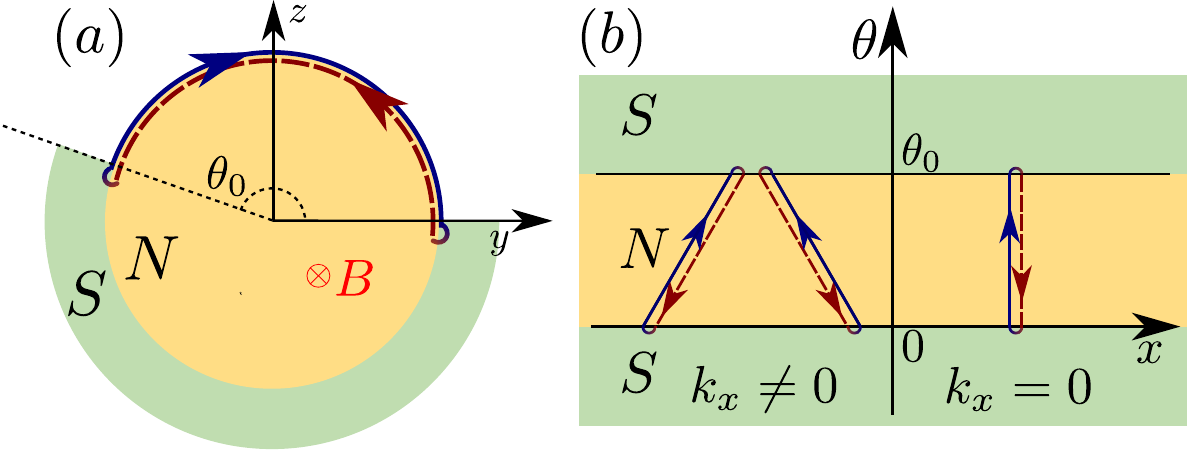}}
\caption{(Color online) (a) A cross section of a NS hybrid junction. The magnetic field is parallel to the wire axis, while the Andreev bound state trajectories are confined to the nanowire surface.
(b) The equivalent two-dimensional system defined on the plane ($x,\theta$).
Since we neglect the possibility for electrons to tunnel through the superconductor, we consider the superconducting leads at $\theta<0$ and $\theta>\theta_0$ infinite.
\label{fig:orb}
}
\end{figure}

\comment{Orbital effect is important, and we analyze it in thin shell limit.}
We now turn to evaluate the consequences of the orbital effect of the magnetic field, known to strongly influence MBS properties~\cite{Lim2013, osca_majorana_2014, Osca2015, Nijholt2016}, in the short-junction limit.
This effect does not manifest in the model of Sec.~\ref{sec:1stmodel} when magnetic field points in the $x$ direction.
To include the orbital effect we use a thin shell approximation, when the electron wave function in the semiconductor is confined to its surface, similar to the system studied in Ref.~\onlinecite{osca_majorana_2014}.
However unlike Ref.~\onlinecite{osca_majorana_2014} we do not assume a constant induced gap and consider instead the nanowire contacted by a bulk superconductor, as shown in Fig.~\ref{fig:orb}.
We model the coupling to the superconductor as two infinite planar superconductors on each side of the 2D uncovered wire section.
By doing so we neglect the possibility for electrons to tunnel through the superconductor to the other side of the uncovered section, which is justified by the density mismatch between the superconductor and the semiconductor.
The thin shell limit is oversimplified and it overestimates the orbital effect of a magnetic field, however it provides an upper bound on the impact of the orbital effect and remains analytically tractable.

The superconductor covers the wire over an angle $2\pi-\theta_0$, while both the wire and the superconductor are translationally invariant in the $x$ direction.
In cylindrical coordinates $(x, \theta)$ the electron Hamiltonian on the nanowire surface reads:
\begin{equation}
H=\left[\frac{p_\theta^2 + p_x^2}{2m}
-\mu\right]\sigma_0-\frac{\alpha}{\hbar}p_x\sigma_y
+E_Z\sigma_x,
\end{equation}
with $p_\theta = -i\hbar R^{-1}\partial/\partial \theta$, and $R$ the radius of the nanowire.
We assume that the magnetic field is fully screened from the superconductor and choose a gauge where the vector potential $A=0$ in the uncovered part of the surface, while the two superconducting leads have a phase difference $\phi = (2 e/\hbar) \pi B R^2$.
Compared to the previous sections where the treatment was more general, we assume from the start the transparent junction limit, when Fermi velocities at $k_x=0$ are identical in the superconductor and the semiconductor and we also neglect the spin-orbit coupling in the transverse direction, as appropriate for $l_\textrm{SO} \gg R \theta_0$.

\begin{figure}[t]
\includegraphics[width=\columnwidth]{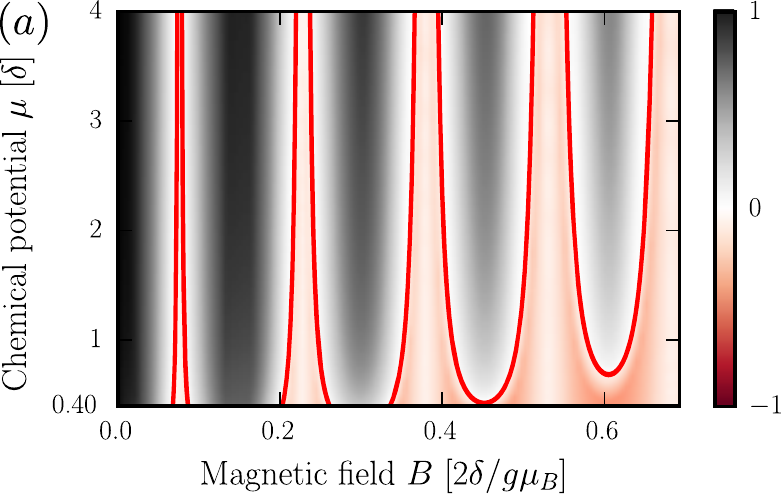}
\centering{\includegraphics[width=\columnwidth]{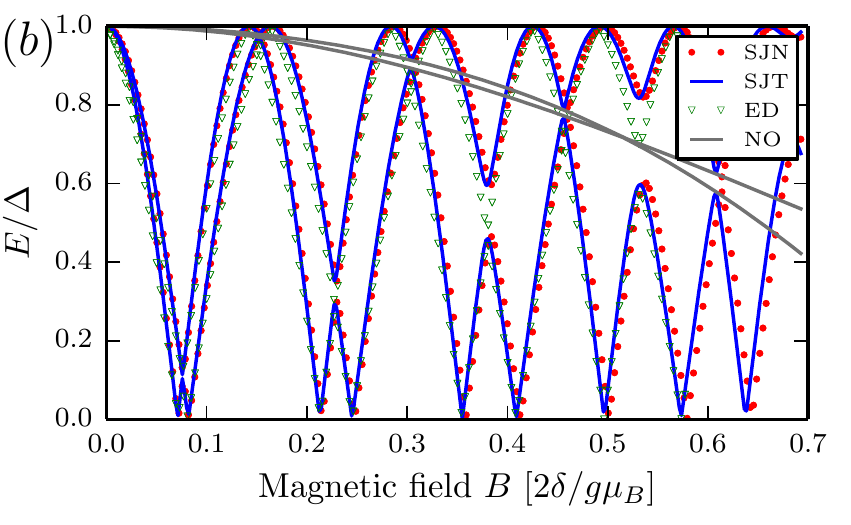}}
\caption{(Color online) (a) Majorana phase diagram $\mathcal{Q} \Delta_\textrm{spec}$ of a transparent NS junction with $\mu_s=\mu_n=\mu$, $m_{s,x}=m_\parallel$, $m_{s,y}=m$, $m_n=m$, and $m_\parallel=10 m$, as a function of chemical potential and magnetic field.
The covering angle is $\theta_0 = \SI{2}{rad}$, so that the width of the uncovered section $R\theta_0$ is equal to the wire diameter.
(b) An example of Andreev spectrum at $k_x=0$, $\mu_n=\mu_s=\SI{3}{\meV}$, and other parameters the same as in (a) in the presence and absence of the orbital effect.
Panel (b) additionally presents a comparison between short junction theoretical (SJT), numerical (SJN), and exact diagonalization (ED).
The theoretical spectrum without orbital effect (NO) is in gray.
The magnetic field is in units of $2\delta/g\mu_B$ with $\delta=\hbar^2\pi^2/2mR^2\theta_0^2$.}
\label{fig:gaps_orb}
\end{figure}

We solve the scattering problem in the basis of conserved spin projections set by Eq.~\eqref{spinors} corresponding to the basis of incoming and outgoing modes:
\begin{equation}
\mathbf a^T=(a^+_L,a^-_L, a^+_R, a^-_R),
\quad\mathbf b^T=(b^+_L,b^-_L, b^+_R, b^-_R),
\end{equation}
with $\mathbf a$ and $\mathbf b$ the amplitudes of incoming and outgoing modes, $R$ denoting the modes at $\theta \leq 0$, $L$ the modes at $\theta \geq \theta_0$, and $\pm$ superscript corresponding to the two conserved spin directions~\eqref{spinors}.
For each spin projection the scattering matrix is given by the classic result for transmission through a potential barrier:
\begin{equation}
S_\pm=\pmat{
r_\pm & t_\pm \\
t_\pm & r_\pm},
\end{equation}
with
\begin{align}
r_\pm &=\frac{(q^2-k^2_\pm)\sin(k_\pm L)}{(q^2+k^2_\pm)
\sin(k_\pm L)+2iqk_\pm\cos(k_\pm L)},
\notag\\
t_\pm &=\frac{2iqk_\pm}{(q^2+k^2_\pm)\sin(k_\pm L)+2iqk_\pm\cos(k_\pm L)},
\end{align}
with momenta $k_\pm$ and $q$ defined by Eq.~\eqref{aniso}.
We then transform the scattering matrix to the basis of time-reversed modes~\eqref{eq:psi_s} and calculate the Andreev spectrum using Eq.~\eqref{nsspec} with the phases of superconducting leads equal to $\phi_R=0$ and $\phi_L = \phi$.
We verify again that the dispersion relation obtained this way agrees well with two numerical tight-binding simulations at fixed chemical potential and that the difference also stays small if we include spin-orbit coupling in the transverse direction [see Fig.~\ref{fig:gaps_orb}(b)].
As before, the spectrum is generically gapped except at $k_x=0$, where topological phase transitions occur.

The resulting Majorana phase diagram is shown in Fig.~\ref{fig:gaps_orb}(a), and it consists of several narrow topological regions centered around $\phi_k = (2k+1)\pi$ with $k$ integer.
At these values of magnetic field, the two superconducting leads have a phase difference of $\pi$, thus fully suppressing the induced gap in the transparent limit.
The Zeeman field then opens a topological gap resulting in a finite extension of the topological phases around $\phi_k$.
We conclude that in the thin shell limit, the orbital effect of the magnetic field reduces $B_*$ by a factor $\sim 10$ for typical junction parameters (we once again note that the thin shell limit overestimates the orbital effect of the magnetic field).
Despite that, it is the Zeeman field responsible for opening the topological gap.

\section{Numerical study of a three-dimensional nanowire}
\label{sec:3d}

\comment{As a final check we study 3D}
To confirm our findings in a model with a more realistic geometry, we numerically calculate the phase diagram of a three-dimensional nanowire in the short-junction limit.
The system consists of a semiconductor nanowire infinite in the $x$ direction and with a square cross section contacted by a bulk superconductor occupying $y<0$ half space  [see Fig.~\ref{fig:1stmodel}(a)].

Due to the large Fermi surface mismatch between the superconductor and the semiconductor we neglect the electron dispersion in the $x$ and $z$ directions in the superconductor.
Therefore, following Sec.~\ref{sec:phys_params} we set $m_{s,y} = m_n= 0.015\, m_e$ and $m_{s,x} = m_{s, z} = \infty$.
Since the semiconductor modes with different values of $k_z$ have different interface transparencies, we cannot ensure a transparent interface for all the modes and instead fix $\mu_s=\SI{8}{\meV}$ while varying $\mu_n\equiv\mu$.
The remaining system parameters are specified in Sec.~\ref{sec:phys_params}.

The model Hamiltonian is a three-dimensional generalization of Eq.~\eqref{danduke} discretized on a cubic lattice.
We include the orbital effect of the magnetic field using Peierls substitution in the gauge $\bm A={\left( 0, 0, B y\Theta(y) \right)}^{T}$.
This ensures that the vector potential is constant in the $x$ direction and that it vanishes at the interface with the superconductor.

\begin{figure}
\includegraphics[width=\columnwidth]{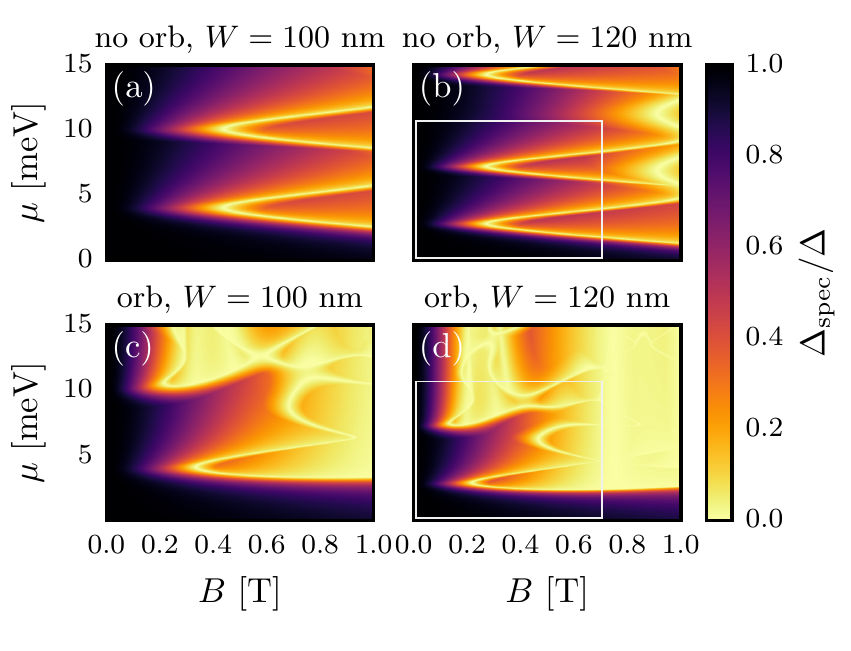}
\caption{(Color online) Spectral gap $\Delta_\mathrm{spec}/\Delta$ dependence on chemical potential $\mu$ in the nanowire and magnetic field $B$ of a square nanowire without (top panels) and with (bottom panels) orbital effect of the magnetic field.
Panels (a), (c) show results for wire section $\SI{100}{\nm} \times \SI{100}{\nm}$, while (b), (d), for $\SI{120}{\nm} \times \SI{120}{\nm}$.
The white box in the right panels shows the same parameter range rescaled by a factor $\left(100/120\right)^2$ to highlight the $W^2$ scaling of the phase diagram.}
\label{fig:phase_diagrams_3d}
\end{figure}

We calculate the excitation spectrum using Eq.~\eqref{nsspec} to find $\Delta_\mathrm{spec}$ by minimization over $k_x$.
The resulting phase diagram of $\Delta_\mathrm{spec}$ is shown in Fig.~\ref{fig:phase_diagrams_3d}.
Comparing the top panels of Fig.~\ref{fig:phase_diagrams_3d} with Fig.~\ref{fig:gaps} we observe the two sharp minima of $B_*(\mu)$.
These correspond to the appearance of the additional bands with a different value of $k_z$ and a minimum in the interface transparency.~\footnote{The code for the tight-binding simulations of the two-dimensional and three-dimensional junctions is available in ancillary files to the manuscript.}

Similar to our observations from Sec.~\ref{sec:orb}, the orbital effect of the magnetic field has a strong effect on the shape of the topological phase boundaries and reduces both $\Delta_\textrm{spec}$ and $B_*$ similar to the thin shell simulations.
Increasing the cross section of the wire [Fig.~\ref{fig:phase_diagrams_3d}(a), (c), against (b), (d)] confirms that in 3D the critical fields preserve the scaling with $B_*\sim 1/W^2$ independent of the presence or absence of orbital effects.

\section{Conclusions and outlook}
\label{sec:conc}

\comment{Superconducting gap is unimportant.}
We have studied the impact of a small superconducting gap on the properties of MBS in semiconductor-superconductor junctions.
The short-junction formalism, appropriate for this limit, allows us to draw universal conclusions about the MBS properties.
Contrary to the intuitive expectations, we show that the reduction of the superconducting gap does not alter the Majorana phase diagram and does not change the size of the MBS.
We therefore conclude that in most practical systems the superconducting gap should not be used as an important parameter in optimizing MBS properties.

\comment{High interface transparency is advantageous.}
On the other hand, we find that the transparency of the semiconductor-superconductor boundary has an important and previously overlooked effect on the Majorana phase diagram.
An interface with $T \approx 1$ produces a phase boundary between trivial and topological phases which depends weakly on the chemical potential.
This is in contrast to $T \ll 1$, used in most prior research, that results in the critical magnetic field having an oscillatory dependence on chemical potential with minima corresponding to the opening of a new band.

\comment{Orbital effect of magnetic field cannot be avoided by superconductor choice or tuning system size.}
Orbital effect of magnetic field plays a dual role: It reduces the critical magnetic field as well as the spectral gap in the topological regime.
Contrary to the predictions of a phenomenological model that assumes a constant induced gap, we show that relative importance of magnetic field cannot be controlled by the superconducting gap or the diameter of the nanowire.

\comment{2DEGs are probably best for Majoranas}
Our findings suggest that creation of MBS in proximitized two-dimensional electron gases laterally contacted by a superconductor is a promising direction of further research.
In these systems the relative strength of the orbital and the Zeeman effect of magnetic field is controlled by an extra tuning parameter: the ratio between the semiconductor thickness and its width.
Additionally, the critical magnetic field in such devices could be tuned using a side gate, effectively changing the semiconductor width without altering the superconductor-semiconductor interface transparency.

\comment{Insensitivity to chemical potential could mean insensitivity to disorder.}
Another important further direction of research is the interplay between junction transparency and disorder.
Since a transparent interface results in a weaker dependence of the critical magnetic field on the chemical potential, it is reasonable to conjecture that the sensitivity of MBS properties to disorder is also reduced in the transparent regime.

\acknowledgements{
The authors thank B.~van Heck, M.~Irfan, M.~P.~Nowak, T.~\"O.~Rosdahl, and M.~Wimmer for many fruitful discussions.
This research was supported by the Foundation for Fundamental Research on Matter (FOM), the Netherlands Organization for Scientific Research (NWO/OCW) as part of the Frontiers of Nanoscience program, and an ERC Starting Grant.}

\appendix
\section{Interface transparency in a two-dimensional junction}
\label{sec:app}
The validity of the short-junction approximation depends on NS interface transparency $T$.
In this section, we review the transparency of a sharp interface between two materials with a parabolic dispersion.
We provide quantitative arguments for the choice of anisotropic mass in the superconductor in modeling a transparent interface.

We consider a planar NS interface with the boundary located at $y=0$, and both materials occupying a half-plane, and solve the scattering problem as outlined in Sec.~\ref{sub:gen}.
Also following Sec.~\ref{sub:gen}, we neglect the spin-orbit scattering at the interface, and use the boundary condition~\eqref{bc}.
At a given energy $E$ there are generally two modes in the semiconductor and two spin-degenerate modes in the superconductor with momenta in the $y$ direction: $k_\pm$ and $q$, respectively:
\begin{eqnarray}
k_\pm &=& \left[\frac{2m_n}{\hbar^2}(E+\mu_n\mp\sqrt{E_Z^2+\alpha^2 k_x^2})
-k_x^2
\right]^{1/2},\\
q &=& \left[\frac{2m_\perp}{\hbar^2}(E+\mu_s)
-\frac{m_\perp}{m_\parallel}k_x^2\right]^{1/2},\notag
\end{eqnarray}
where we use the same notation as in Sec.~\ref{sub:gen} the superconductor has anisotropic mass $(m_\perp, m_\parallel)$.

\begin{figure}[t]
\includegraphics[width=\columnwidth]{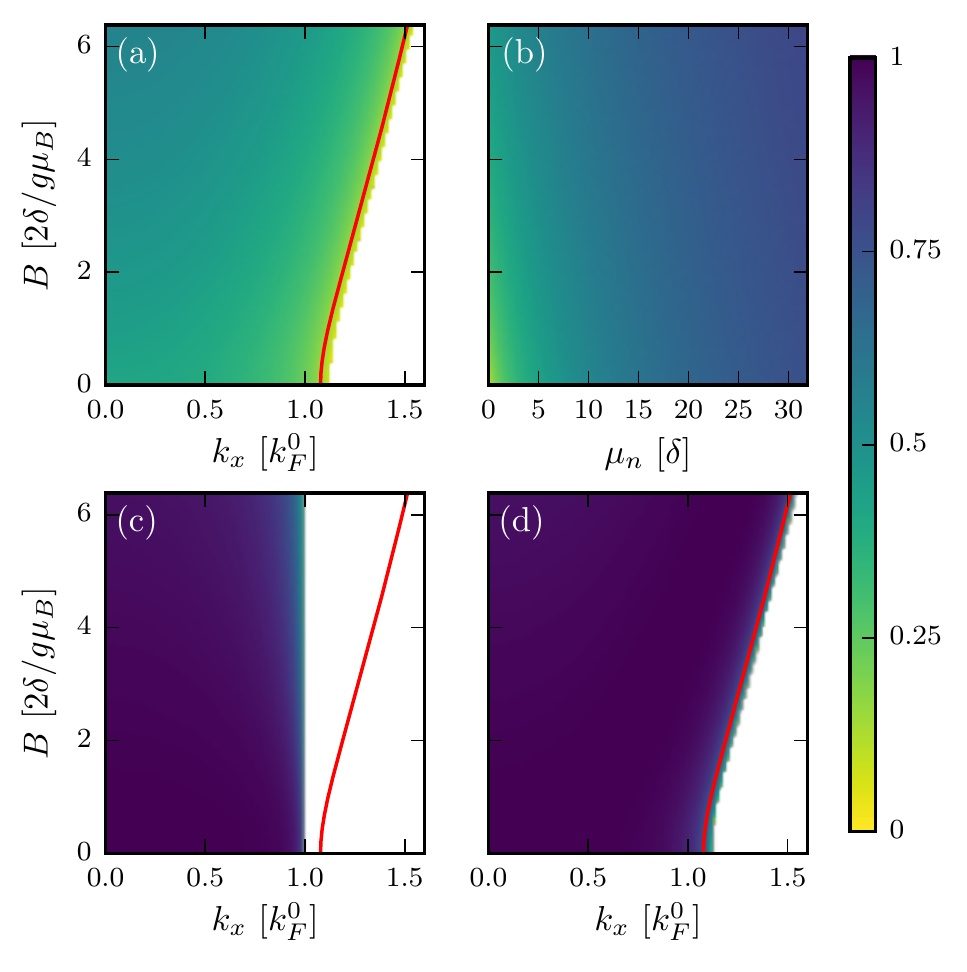}
\caption{(Color online)
Transmission probability of one of the spin polarizations $T_-$ of an infinite NS junction.
(a, c, and d) $T_-$ as a function of magnetic field $B$ in units of $\frac{2\delta}{g\mu_B}$, with $\delta$ defined in Eq.~\eqref{eq:e_thouless} and parallel momentum $k_x$ (in units of $k_F^0=\sqrt{2m_n\mu_n/\hbar^2}$).
The momenta $k_x$ run over the Fermi surface of the semiconductor, which is marked by the red line.
(a) Using bare material parameters $\mu_n=\SI{3}{\meV}$, $\mu_s=\SI{11.7}{eV}$, $m_s=m_e$, $m_n=0.015\;m_e$.
(b) $T_-$ versus $B$ and $\mu_n$ at $k_x=0$ and the same parameters as in (a).
(c) $T_-$ when the chemical potential and mass are equal in the superconductor and the semiconductor.
(d) $T_-$ for anisotropic mass in the superconductor $(m_\parallel, m)$, $m_n\equiv m=0.015\;m_e$, with $m_\parallel=10\;m$, $\mu_n=\mu_s=\SI{3}{\meV}$.
Only evanescent solutions exist in the white regions; transmission $T_-$ becomes imaginary.
}
\label{fig:transp}
\end{figure}

The transmission probabilities of two spin orientations $(\pm)$ follow immediately:
\begin{equation}
T_\pm=4
\bigg(\sqrt\frac{v_\pm}{v_s}+\sqrt\frac{v_s}{v_\pm}\bigg)^{-2},
\end{equation}
with $v_s=\hbar q/m_\perp$ and $v_\pm=\hbar k_\pm/m_n$ the velocities normal to the interface in two materials.
Both $T_+$ and $T_-$ exhibit similar behavior, except for $T_+$ vanishing inside the helical gap.
In contrast, $T_-$ is always well defined at $k_x=0$ for $\mu_n>0$.
For concreteness, we illustrate the dependence of $T_-$ on the Hamiltonian parameters.

Let us first start with realistic parameters both in the superconductor and the semiconductor.
The chemical potential in the nanowire $\mu_n$ is gate tunable.
We choose to fix it at \SI{3}{\meV}, comparable to the level spacing in a nanowire.
The rest of the parameters are specified in Sec.~\ref{sec:phys_params}.
The results are plotted in Fig.~\ref{fig:transp}(a), for all momenta in the semiconductor Fermi surface and an experimentally relevant range of magnetic fields.
The transparency is mostly around $40\%$ but rapidly vanishes near the Fermi momentum.
Modifying the chemical potential in the wire does not appreciably increase the transparency [see Fig.~\ref{fig:transp}(b)].
The low transparency is artificial and due to the choice of a sharp change in mass and chemical potential across the interface.

Choosing $m_\perp = m_\parallel = m_n$ and $\mu_s = \mu_n$ results in a nearly perfect transmission at all angles, as shown in Fig.~\ref{fig:transp}(c).
However, this parameter choice is also unphysical since the semiconductor Fermi surface becomes larger than the superconductor one at any finite magnetic field.
Then the interface becomes opaque for higher momenta $k_x$.

Finally, choosing $\mu_s = \mu_n$ and an anisotropic mass in the superconductor $(m_{s,x},m_{s,y})=(m_\parallel, m_n)$, with $m_\parallel \gg m_n$ results in an interface that stays transparent for all $k_x$ [see Fig.~\ref{fig:transp}(d)].

\section{Independence of critical magnetic field on the superconducting gap}
\label{sec:app2}

\begin{figure}[h]
\includegraphics[width=0.9\columnwidth]{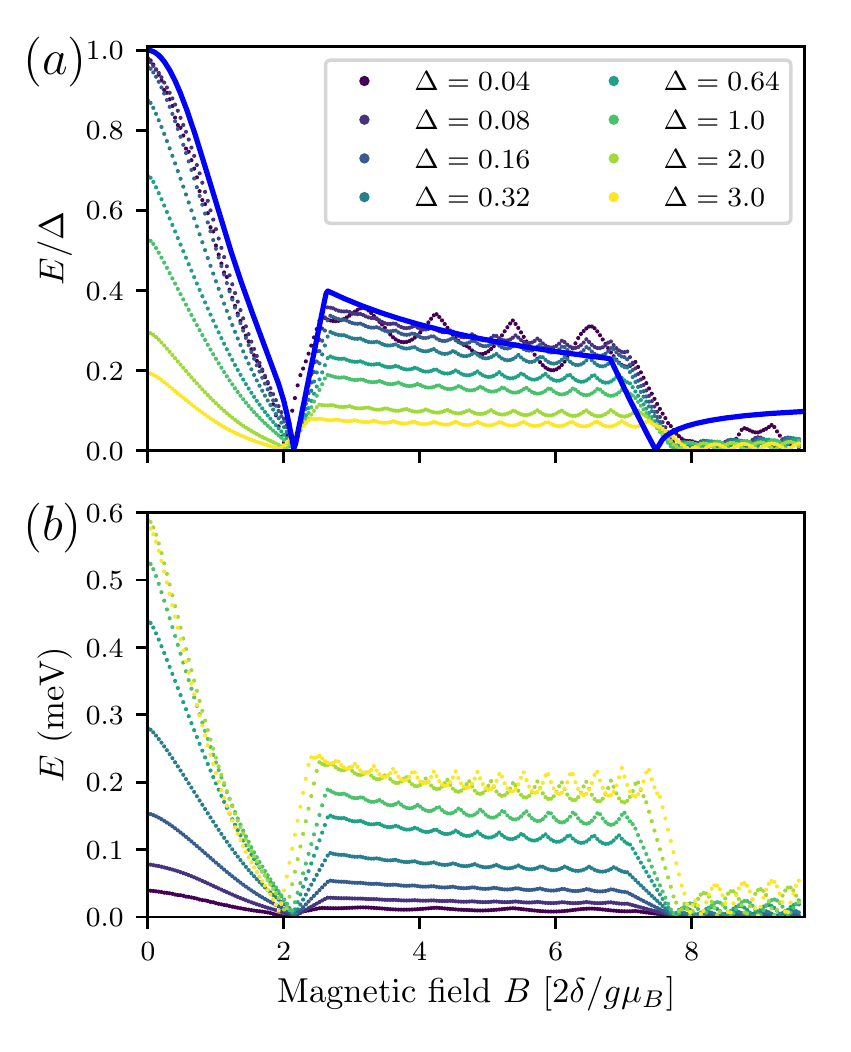}
\caption{(Color online) The energy of the second excited state of a finite nanowire junction calculated using exact diagonalization.
Up to the second topological phase transition it is a good approximation of the induced gap.
The eigenenergies are normalized to $\Delta$ in panel (a) and unnormalized in panel (b).
In the long junction regime the induced gap tends to a constant, while in the short-junction regime the ratio $E/\Delta$ tends to the analytical result derived for the short-junction limit.
The legend applies to both panels.
The superconducting gaps $\Delta$ are in meV.}
\label{fig:scaling_delta}
\end{figure}

Here we verify the validity of our conclusions about the scaling of the eigenenergies and the independence of $B_*$ on $\Delta$ using exact diagonalization of a finite BdG Hamiltonian for the semiconductor-superconductor heterostructure at different values of the superconducting gap.
We use the same setup of a finite heterojunction modeled with the BdG Hamiltonian~\eqref{sc} as in Sec.~\ref{sec:finite}.
We check the behavior of the critical field and the bulk band gap by tracking the energy of the second excited state while varying $B$ for values of $\Delta$ ranging from $\SI{40}{\mu\eV}$ to $\SI{3}{\meV}$ (almost two orders of magnitude).
Our results for a heterojunction of size $\SI{3000}{\nm}\times\SI{6000}{\nm}$ with a normal region occupying $\SI{3000}{\nm}\times\SI{100}{\nm}$ are shown in Fig.~\ref{fig:scaling_delta}.

When $\Delta \gtrsim E_\mathrm{Th}\approx \SI{1}{\meV}$ the system transitions to the long junction regime, so that the ratio $\Delta_\textrm{spec}/\Delta$ continues to decrease, while $\Delta_\text{spec}$ becomes almost independent on $\Delta$.
In the opposite limit $\Delta \ll E_\mathrm{Th}$, we observe that $\Delta_\textrm{spec}/\Delta$ tends to a constant, in agreement with the short-junction limit prediction.
The field values $B_*$ where $\Delta_\textrm{spec}$ vanishes stay almost constant, with the residual variation due to the effect of a finite system size and lattice constant.

\bibliographystyle{apsrev4-1}
\bibliography{shortjunction}
\end{document}